\documentclass[prd,a4paper,superscriptaddress,nofootinbib,showpacs,
amssymb]{revtex4}
\usepackage{graphics}
\usepackage{epsfig}
\usepackage{amsmath}

\newcommand{\g}{\gamma}

\def\l{\left}
\def\r{\right}

\def\mat#1  { \begin{matrix}#1\end{matrix} }
\def\pmat#1 { \begin{pmatrix}#1\end{pmatrix} }
\def\cas#1  { \begin{cases}#1\end{cases} }

\begin{document}

\title{Braneworld tensor anisotropies in the CMB}

\author{Bernard Leong}
\email{cwbl2@mrao.cam.ac.uk} \affiliation{Astrophysics Group,
Cavendish Laboratory, Madingley Road, Cambridge CB3 OHE, UK}
\author{Anthony Challinor}
\email{a.d.challinor@mrao.cam.ac.uk} \affiliation{Astrophysics
Group, Cavendish Laboratory, Madingley Road, Cambridge CB3 OHE,
UK}
\author{Roy Maartens}
\email{roy.maartens@port.ac.uk} \affiliation{Institute of
Cosmology \& Gravitation, Portsmouth University, Portsmouth PO1
2EG, UK}
\author{Anthony Lasenby}
\email{a.n.lasenby@mrao.cam.ac.uk} \affiliation{Astrophysics
Group, Cavendish Laboratory, Madingley Road, Cambridge CB3 OHE,
UK}
\date{\today}

\begin{abstract}
Cosmic microwave background (CMB) observations provide in
principle a high-precision test of models which are motivated by M
theory. We set out the framework of a program to compute the
tensor anisotropies in the CMB that are generated in braneworld
models. In the simplest approximation, we show the braneworld
imprint as a correction to the power spectra for standard temperature and
polarization anisotropies.
\end{abstract}
\maketitle

\section{Introduction}

The early universe provides a testing ground for theories of
gravity. The standard cosmological model, based on general
relativity with an inflationary era, is very effective in
accounting for a broad range of observed features of the universe.
However, the lack of a consistent theoretical framework for
inflation, together with the ongoing puzzles on the nature of dark
matter and dark energy, indicate that cosmology may be probing the
limits of validity of general relativity.

M theory is considered to be a promising potential path to quantum
gravity. As such, it is an important candidate for cosmological
testing. In the absence of a sufficiently general M-theoretic
model of cosmology, we can use phenomenological models that share
some of the key features of M theory, including branes. In brane
cosmology, the observable universe is a 1+3-dimensional ``brane"
surface moving in a higher-dimensional ``bulk" spacetime.
Standard-model fields are confined to the brane, while gravity
propagates in the bulk. The simplest, and yet sufficiently
general, phenomenological braneworld models are those based on the
Randall-Sundrum~II scenario~\cite{randall}. These models have the
additional advantage that they provide a framework for
investigating aspects of holography and the AdS/CFT
correspondence.

In the generalized RSII models, analyzed via an elegant
geometrical approach in Ref.~\cite{shiromizu}, the bulk is a
1+4-dimensional spacetime, with non-compact extra spatial
dimension. What prevents gravity from `leaking' into the extra
dimension at low energies is the negative bulk cosmological
constant $\Lambda_5=-6/\ell^2$, where $\ell$ is a curvature scale
of the bulk. In the weak-field static limit, null results in tests
for deviations from Newton's law impose the limit $\ell \lesssim
1$~mm. The negative $\Lambda_5$ is offset by the positive brane
tension $\lambda$, which defines the energy scale dividing low
from high energies. The limit $\ell<1~$mm implies
$\lambda>(100~{\rm GeV})^4$, and the effective cosmological
constant on the brane is
\begin{equation}\label{cc}
\Lambda=\frac{1}{2} (\Lambda_5+\kappa^2\lambda)\,,
\end{equation}
where $\kappa^2= 8\pi G=8\pi/M_4^2$, and $M_4\sim 10^{19}~$GeV is
the effective Planck scale on the brane. A further intriguing
feature of the braneworld scenario is that, because of the large
extra dimensions, the fundamental energy scale of gravity can be
dramatically lower than the effective Planck scale on the brane --
as low as $\sim$~TeV in some scenarios. In generalized RSII
models, the fundamental scale is higher, $M_5>10^5~$TeV, and is
related to $M_4$ via $M_5^3=M_4^2/\ell\,.$

At energies well above the brane tension $\lambda$, gravity
becomes 5-dimensional and significant corrections to general
relativity occur. There are also corrections that can operate at
low energies, mediated by bulk graviton or Kaluza-Klein (KK)
modes. Both types of correction play an important role in tensor
perturbations.

The background cosmological dynamics of a Friedmann brane in
Schwarzschild-Anti de Sitter (AdS) bulk are well
understood~\cite{binetruy}, including the high-energy
modifications to inflation~\cite{maartens3}. High-energy inflation
on the brane generates a zero-mode (4D graviton mode) of tensor
perturbations, and stretches it to super-Hubble scales. This
zero-mode has the same qualitative features as in general
relativity, remaining frozen at constant amplitude while beyond
the Hubble horizon, but the overall amplitude is
higher~\cite{langlois5}. The massive KK modes (5D graviton modes)
remain in the vacuum state during slow-roll inflation. The
evolution of the super-Hubble zero mode is the same as in general
relativity, so that high-energy braneworld effects in the early
universe serve only to re-scale the amplitude. However, when the
zero mode re-enters the Hubble horizon, massive KK modes can be
excited. Qualitative arguments~\cite{hawking,gorbunov1} indicate
that this is a very small effect, but it remains to be properly
quantified, so that the signature on the CMB may be calculated,
and constraints may be imposed on the braneworld parameters.

We develop here a formalism to compute the tensor anisotropies in
the CMB, which incorporates the early-universe high-energy
braneworld effects, and we carefully delineate what is known on
the brane from what is required from bulk equations. Once the 5D
solutions are provided, our formalism, with its modified CMB code
(based on CAMB~\cite{lewis,lewis1}), is able to compute these
anisotropies. We illustrate this by using a simple approximation
to the 5D effects (cf. the analysis of the braneworld scalar
Sachs-Wolfe effect in Ref.~\cite{barrow}).

\section{Braneworld dynamics and tensor perturbations}

There has been an explosion of interest in the theory of
cosmological perturbations in braneworlds and their implications
for observational cosmology (see e.g. Refs.
\cite{pert}).

In this paper, we follow the same general formalism previously
developed to analyze braneworld scalar anisotropies in the
CMB~\cite{leong1}. This is based on the 1+3-covariant perturbation
theory for CMB anisotropies~\cite{challinor4}, generalized to
incorporate braneworld effects. (See Ref.~\cite{leong1} for
further details and definitions.)

The field equations induced on the brane are
\begin{equation} \label{e:einstein1}
G_{ab} = - \Lambda g_{ab} + \kappa^2 T_{ab} +
6\frac{\kappa^2}{\lambda} {\cal S}_{ab} - {\cal E}_{ab}\;,
\end{equation}
where ${\cal S}_{ab}\sim (T_{ab})^2$ carries high-energy
corrections, and is negligible for matter energy densities
$\rho\ll\lambda$, while ${\cal E}_{ab}$, the projection of the
bulk Weyl tensor on the brane, carries corrections from KK or 5D
graviton effects. From the brane-observer viewpoint, the
energy-momentum corrections in ${\cal S}_{ab}$ are local, whereas
the KK corrections in ${\cal E}_{ab}$ are
nonlocal~\cite{maartens,mukohyama}, since they incorporate 5D
gravity wave modes. These nonlocal corrections to the Einstein
equations on the brane cannot be determined purely from data on
the brane, and so the induced field equations are not a closed
system; one needs to supplement them by 5D equations governing
${\cal E}_{ab}$.

The trace free ${\cal E}_{ab}$ contributes an effective energy
density $\rho^*$, pressure $\rho^*/3$, momentum density $q^*_a$
and anisotropic stress $\pi^*_{ab}$ on the brane, which
collectively incorporate the spin-0, spin-1 and spin-2 modes of
the 5D graviton, and through which the brane ``feels" the bulk
gravitational field. Then the braneworld corrections can
conveniently be consolidated into an effective total energy
density, pressure, momentum density and anisotropic stress:
\begin{eqnarray}
\rho^{\text{eff}} &=& \rho\left(1 +\frac{\rho}{2\lambda} +
\frac{\rho^*}{\rho} \right)\;, \\ \label{e:pressure1} P^{\text{eff
}} &=& P  + \frac{\rho}{2\lambda} (2P+\rho)+\frac{\rho^*}{3}\;, \\
q^{\text{eff }}_a &=& q_a\left(1+  \frac{\rho}{\lambda} \right)
+q^*_a \\ \label{e:pressure2} \pi^{\text{eff }}_{ab} &=& \pi_{ab}
\left(1-\frac{\rho+3P}{2\lambda}\right)+\pi^*_{ab}\;,
\end{eqnarray}
where $\rho$ and $P$ are the total matter density and pressure,
$q_a$ is the total matter momentum density, and $\pi_{ab}$ is the
matter anisotropic stress. We have neglected terms quadratic in
$q_a$ and $\pi_{ab}$ since these do not contribute to the
fluctuations in the effective variables in linear perturbation
theory.

In the background, $q^*_a=0=\pi^*_{ab}$. The modified Friedmann
equations are
\begin{eqnarray}
H^2 &=& \frac{\kappa^2}{3} \rho^{\text{eff }} + \frac{1}{3}
\Lambda - \frac{K}{a^2} \,, \\ \label{e:friedmann1} \dot H &=&
-\frac{\kappa^2}{2}(\rho^{\rm eff} +P^{\rm eff})+\frac{K}{a^2}\,,
\end{eqnarray}
and the KK energy density behaves like `dark' radiation:
\begin{equation}
\rho^* \propto\frac{1}{a^4}\,.
\end{equation}

The (1+3)-covariant description of tensor perturbations in general
relativity~\cite{dunsby4,challinor4} generalizes naturally to
braneworld cosmology~\cite{maartens}. Tensor modes on the brane
are characterized by the transverse traceless shear $\sigma_{ab}$,
which is a tensor potential for the electric and magnetic Weyl
tensors $E_{ab}$ and $H_{ab}$. The sources for linearized tensor
modes are the matter anisotropic stress $\pi_{ab}$ and the KK
anisotropic stress $\pi^*_{ab}$. The Boltzmann equation governs
$\pi_{ab}$. As pointed out above, the KK anisotropic stress, which
carries the imprint of 5D gravitational wave modes on the brane,
is not determined by 4D equations on the brane. For the moment, we
assume that $\pi^*_{ab}$ is given from a solution to the 5D
equations. Below we will discuss what this entails and how one can
approximate the solution.

The transverse traceless quantities can be expanded in electric
($Q_{ab}^{(k)} $) and magnetic ($\bar{Q}_{ab}^{(k)} $) parity
tensor harmonics~\cite{challinor4}, with dimensionless
coefficients:
\begin{align}
E_{ab} &= \sum_{k} \l(\frac{k}{a}\r)^2 \left[E_k Q_{ab}^{(k)} +
\bar{E}_k \bar{Q}_{ab}^{(k)} \right], \\ H_{ab} &= \sum_{k}
\left(\frac{k}{a}\r)^2 \l[H_k Q_{ab}^{(k)} + \bar{H}_k
\bar{Q}_{ab}^{(k)} \right], \\ \sigma_{ab} &= \sum_{k} \frac{k}{a}
\l[\sigma_k Q_{ab}^{(k)} + \bar{\sigma}_k \bar{Q}_{ab}^{(k)}
\right],  \\ \pi_{ab} &= \rho \sum_{k} \left[ \pi_{k} Q_{ab}^{(k)}
+ \bar{\pi}_k \bar{Q}_{ab}^{(k)} \right], \\ \pi_{ab}^{*} &= \rho
\sum_{k} \left[ \pi_{k}^{*} Q_{ab}^{(k)} + \bar{\pi}_k^{*}
\bar{Q}_{ab}^{(k)} \right].
\end{align}
Using $H_{ab}={\rm curl}\,\sigma_{ab}$, we arrive at the coupled
equations
\begin{eqnarray} \label{e:sigmadot}
&& \frac{k}{a^2} \l(\sigma'_k + {\cal H} \sigma_k \r) +
\frac{k^2}{a^2} E_k - \frac{\kappa^2}{2} \rho \pi_k = \kappa^2
(2-3\gamma)\frac{\rho^2}{4\lambda}\pi_k  + \frac{\kappa^2}{2} \rho
\pi_k^{*}\,,\\ && \label{e:Edot} \frac{k^2}{a^2} \l(E'_k + {\cal
H} E_k \r) - k\l(\frac{k^2}{a^2} + \frac{3K}{a^2} -
 \frac{\kappa^2}{2} \g \rho  \r) \sigma_k +
\frac{\kappa^2}{2} \rho \pi'_k -\frac{\kappa^2}{2} (3 \g -1) {\cal
H}\rho \pi_k \nonumber\\ &&{}~= -\frac{\kappa^2}{4\lambda} \{ 2 k
\g \rho^2 \sigma_k - (3 \g-2) \rho^2 \pi'_k-[3 \g' - (3
\g-2)(6\g-1) {\cal H}] \rho^2 \pi_k \} \nonumber\\ &&{}~~  -
\frac{2}{3} k\kappa^2\rho^* \sigma_k  - \frac{\kappa^2}{2} \l[\rho
 \pi_k^{*'} + (1-3\g) {\cal H} \rho \pi_k^{*} \r]\,,
\end{eqnarray}
where a prime denotes $d/d\tau$, with $\tau$ conformal time,
${\cal H}=a'/a$, and the (non-constant) parameter $\gamma$ is
defined by $P=(\gamma-1)\rho$. Equations~$\eqref{e:sigmadot}$ and
$\eqref{e:Edot}$, with all braneworld terms on the right-hand
sides, determine the tensor anisotropies in the CMB, once $\pi_k$
and $\pi^*_k$ are given. The former is determined by the Boltzmann
equation in the usual way~\cite{challinor4}, except that the
background dynamics are altered by braneworld effects. The latter
requires a solution of the 5D perturbation equations.

The solution for the non-local anisotropic stress will be of the
form
\begin{equation}\label{e:soln}
\pi^*_k(\tau) \propto \int d \tilde\tau\,\,{\cal
G}(\tau,\tilde\tau) F[\pi_k,\sigma_k]\big|_{\tilde\tau} \,,
\end{equation}
where ${\cal G}$ is a retarded Green's function evaluated on the
brane. The functional $F$ is known in the case of a Minkowski
background~\cite{sasaki}, but not in the cosmological case. (An
equivalent integro-differential formulation of the problem is
given in Ref.~\cite{mukohyama2}; see also Ref.~\cite{soda}.) Once
${\cal G}$ and $F$ are determined, Eq.~$\eqref{e:soln}$ can in
principle be incorporated into a modified version of Boltzmann
codes such as CAMB~\cite{lewis} or CMBFAST~\cite{cmbfast}. It
remains a major task of braneworld cosmological perturbation
theory to find this solution, or its equivalent forms in other
formalisms. In the meanwhile, in order to make progress towards
understanding braneworld signatures on tensor CMB anisotropies, we
can consider approximations to the solution.

The nonlocal nature of $\pi^*_k$, as reflected in
Eq.~$\eqref{e:soln}$, is fundamental, but is also the source of
the great complexity of the problem. The lowest level
approximation to $\pi^*_k$ is local. Despite removing the key
aspect of the KK anisotropic stress, we can get a feel for its
influence on the CMB if we capture at least part of its
qualitative properties. The key qualitative feature is that
inhomogeneity and anisotropy on the brane are a source for KK
modes in the bulk which ``backreact"~\cite{maartens} or ``feed
back"~\cite{mukohyama}, onto the brane. The transverse traceless
part of inhomogeneity and anisotropy on the brane is given by the
transverse traceless anisotropic stresses in the geometry, i.e. by
the matter anisotropic stress $\pi_{ab}$ and the shear anisotropy
$\sigma_{ab}$. The radiation and neutrino anisotropic stresses are
in turn sourced by the shear to lowest order (neglecting the role
of the octupole and higher Legendre moments).

Thus the simplest local approximation which reflects the essential
qualitative feature of the spin-2 KK modes is
\begin{equation} \label{e:Pansatz1}
\kappa^2 \pi_{ab}^{*} = - \zeta H \sigma_{ab}\,,~~\zeta'=0\,,
\end{equation}
where $\zeta$ is a dimensionless KK parameter, with $\zeta=0$
corresponding to no KK effects on the brane, and
$\zeta=0=\lambda^{-1}$ giving the general relativity limit. [Note
that for tensor perturbations, where there is no freedom over the
choice of frame (i.e.\ $u^a$), there is no gauge ambiguity in
Eq.~(\ref{e:Pansatz1}). However, for scalar or vector
perturbations, this relation could only hold in one frame, since
$\pi_{ab}^{*}$ is frame-invariant in linear theory while
$\sigma_{ab}$ is not.]

The approximation in Eq.~(\ref{e:Pansatz1}) has the qualitative
form of a shear viscosity, which suggests that KK effects lead to
a damping of tensor anisotropies. This is indeed consistent with
the conversion of part of the zero-mode at Hubble re-entry into
massive KK modes~\cite{langlois5,gorbunov1}. The conversion may be
understood equivalently as the emission of KK gravitons into the
bulk, and leads to a loss of energy in the 4D graviton modes on
the brane, i.e. to an effective damping. The approximation in
Eq.~(\ref{e:Pansatz1}) therefore also incorporates this key
feature qualitatively.

With this first approximation, we can close the system of
equations on the brane by adding the equation
\begin{equation}
\kappa^2 \rho \pi_{k}^{*} = - \zeta {\cal H} \frac{k}{a^2}
\sigma_k\,.
\end{equation}
We will also assume $K=0=\rho^*$ in the background. The parameter
$\zeta$ (together with the brane tension $\lambda$) then controls
braneworld effects on the tensor CMB anisotropies in this simplest
approximation.

\section{ CMB Tensor Power Spectra}

Ignoring the photon anisotropic stress (i.e. $\pi_{ab}=0$), the
variable $u_k \equiv a^{1+{\zeta}/{2}} \sigma_k$ satisfies the
equation of motion
\begin{equation} \label{e:ueqn2}
u_k'' + \l[k^2 +2 K - \frac{(a^{-1-
{\zeta}/{2}})''}{a^{-1-{\zeta}/{2}}} \r] u_k = 0\, ,
\end{equation}
where we have used Eq.~(\ref{e:Pansatz1}). In flat models ($K=0$)
on large scales there is a decaying solution $\sigma_k \propto
a^{-(2+\zeta)} $. Since Eq.~(\ref{e:ueqn2}) contains no first
derivative term the Wronskian is conserved. On large scales we can
use the solution $\sigma_k \propto a^{-(2+\zeta)} $ to write the
conserved Wronskian as $W = \sigma_k' + (2+\zeta){\cal H} \sigma_k
$. (The Wronskian vanishes in the decaying mode.) Integrating
gives the following two independent solutions on large scales in
flat models:
\begin{equation}
\sigma_k=
\begin{cases}
A_k a^{-(2+\zeta)} \,,\\ B_k a^{-(2+\zeta)} \int^{\tau}
d\tilde\tau\, a(\tilde\tau)^{2+\zeta} \, ,
\end{cases}
\end{equation}
where $A_k$ and $B_k$ are constants of integration. If we let
$\zeta \to 0$, we recover the results in Ref.~\cite{lewis1} for
the general relativity case.

The conserved Wronskian is proportional to the metric perturbation
variable, $H_T$, characterising the amplitude of 4D gravitational
waves. In flat models $H_T$ is related to the covariant variables
quite generally by
\begin{equation} \label{e:HT1}
H_{Tk} = \frac{\sigma'_k}{k} + 2 E_k \, .
\end{equation}
Ignoring photon anisotropic stress, we can eliminate the electric
part of the Weyl tensor via the shear propagation
equation~(\ref{e:sigmadot}) to find $k H_{Tk} = - \sigma_k' -
(\zeta + 2) \mathcal{H} \sigma_k = -W$. The fact that $H_T$ is
conserved on large scales in flat models in the absence of photon
anisotropic stress can also be seen directly from its propagation
equation,
\begin{equation}
H_{Tk}'' + (2+\zeta) {\cal H} H_{Tk}' + k^2 H_{Tk} = 0\,.
\end{equation}

We can solve Eq.~(\ref{e:ueqn2}) on all scales in the high-energy
($\rho\gg \lambda$ and $a\propto \tau^{1/3}$) and low-energy
($\rho\ll\lambda$ and $a\propto \tau$) radiation-dominated
regimes, and during matter-domination ($a\propto \tau^2$). The
solutions are
\begin{align}
\label{e:highenergy1} u_k(\tau) &= \sqrt{k\tau} \left[ c_1
J_{\frac{1}{6}(5+\zeta)} (k \tau)  + c_2 Y_{\frac{1}{6}(5+\zeta)}
(k \tau)\right] & \quad &(\text{high energy radiation}), \\
\label{e:lowenergy1} u_k(\tau) &= \sqrt{k\tau} \left[c_3
J_{\frac{1}{2} (3+\zeta)} (k \tau) + c_4 Y_{\frac{1}{2}(3+\zeta)}
(k \tau)\right] & &(\text{low energy radiation}),\\
\label{e:matter} u_k(\tau) &= \sqrt{k\tau} \left[ c_5
J_{\frac{5}{2}+\zeta} (k \tau)  + c_6 Y_{\frac{5}{2}+\zeta} (k
\tau)\right] & &(\text{matter domination}),
\end{align}
where $c_i$ are integration constants. The solutions for the
electric part of the Weyl tensor can be found from
Eq.~(\ref{e:sigmadot}). For modes of cosmological interest the
wavelength is well outside the Hubble radius at the transition
from the high-energy regime to the low-energy. It follows that the
regular solution (labelled by $c_1$) in the high-energy regime
will only excite the regular solution ($c_3$) in the low-energy,
radiation-dominated era. Performing a series expansion, we arrive
at the appropriate initial conditions for large-scale modes in the
low-energy radiation era:
\begin{align}
\label{e:hijini} H_{Tk} &=  1 - \frac{(k \tau)^2}{2 (3 + \zeta)} +
\frac{(k \tau)^4}{8 (3+\zeta)(5+\zeta)}  + O[(k\tau)^6], \\
\sigma_k &= -\frac{k \tau}{3+\zeta} + \frac{k^3
\tau^3}{2(3+\zeta)(5+\zeta)} + O[(k\tau)^5], \label{e:shearini}\\
\label{e:elecini} E_k &=  \frac{(4 + \zeta)}{2(3+\zeta)} -
\frac{(k\tau)^2 (8+\zeta)}{4(3+\zeta)(5+ \zeta)} + O[(k\tau)^4].
\end{align}
In the limit $\zeta \to 0$, we recover the general relativity
results~\cite{challinor4}.

For modes that are super-Hubble at matter-radiation equality
(i.e.\ $k \tau_{\text{eq}} \ll 1$), the above solution joins
smoothly onto the regular solution labelled by $c_5$ in
Eq.~(\ref{e:matter}). For $k \tau_{\text{eq}} \gg 1$, the shear
during matter domination takes the form
\begin{equation}
\label{e:matterlong} \sigma_k = -
2^{\frac{3}{2}+\zeta}\Gamma(\tfrac{5}{2}+\zeta)
(k\tau)^{-(\frac{3}{2}+\zeta)} J_{\frac{5}{2}+\zeta}(k\tau).
\end{equation}
In the opposite limit, the wavelength is well inside the Hubble
radius at matter-radiation equality. The asymptotic form of the
shear in matter domination is then
\begin{equation}
\label{e:mattershort} \sigma_k \sim
\frac{\Gamma[\frac{1}{2}(3+\zeta)]}{\sqrt{\pi}} \left(
\frac{2\tau_{\text{eq}}}{\tau}\right)^{1+\zeta/2}
(k\tau)^{-(1+\zeta/2)} \sin(k\tau - \pi\zeta/4).
\end{equation}

\begin{figure}[!bth]
\begin{center}
\includegraphics[scale=0.6]{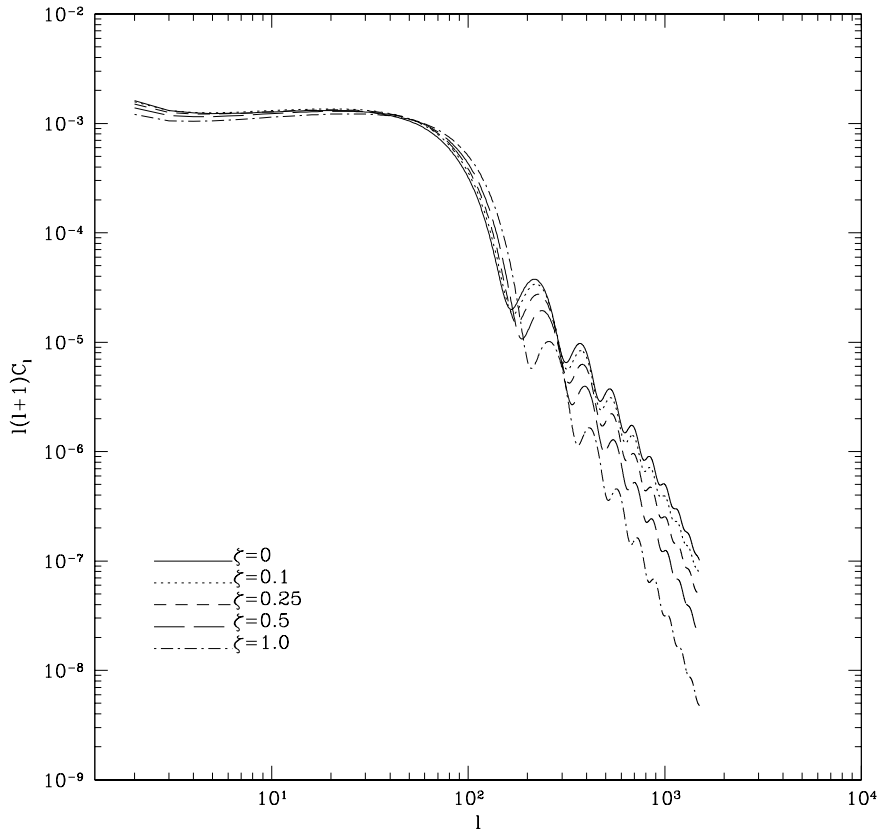}
\caption{The temperature power spectrum for tensor perturbations
in braneworld models using the approximation in
Eq.~(\ref{e:Pansatz1}), with $\zeta$ the dimensionless KK
parameter. Models are shown with $\zeta= 0.0$,  0.1, 0.25,  0.5
and 1.0. The initial tensor power spectrum is scale invariant and
we have adopted an absolute normalisation to the power in the
primordial gravity wave background. The background cosmology is
the spatially flat $\Lambda$CDM (concordance) model with density
parameters $\Omega_b=0.035$,  $\Omega_{c}=0.315$,
$\Omega_{\Lambda}=0.65$, no massive neutrinos, and the Hubble
constant $H_0=65\, \text{km~s}^{-1} \text{Mpc}^{-1}$.
\label{plot2b}}
\end{center}
\end{figure}

\begin{figure}[!bth]
\begin{center}
\includegraphics[scale=0.6]{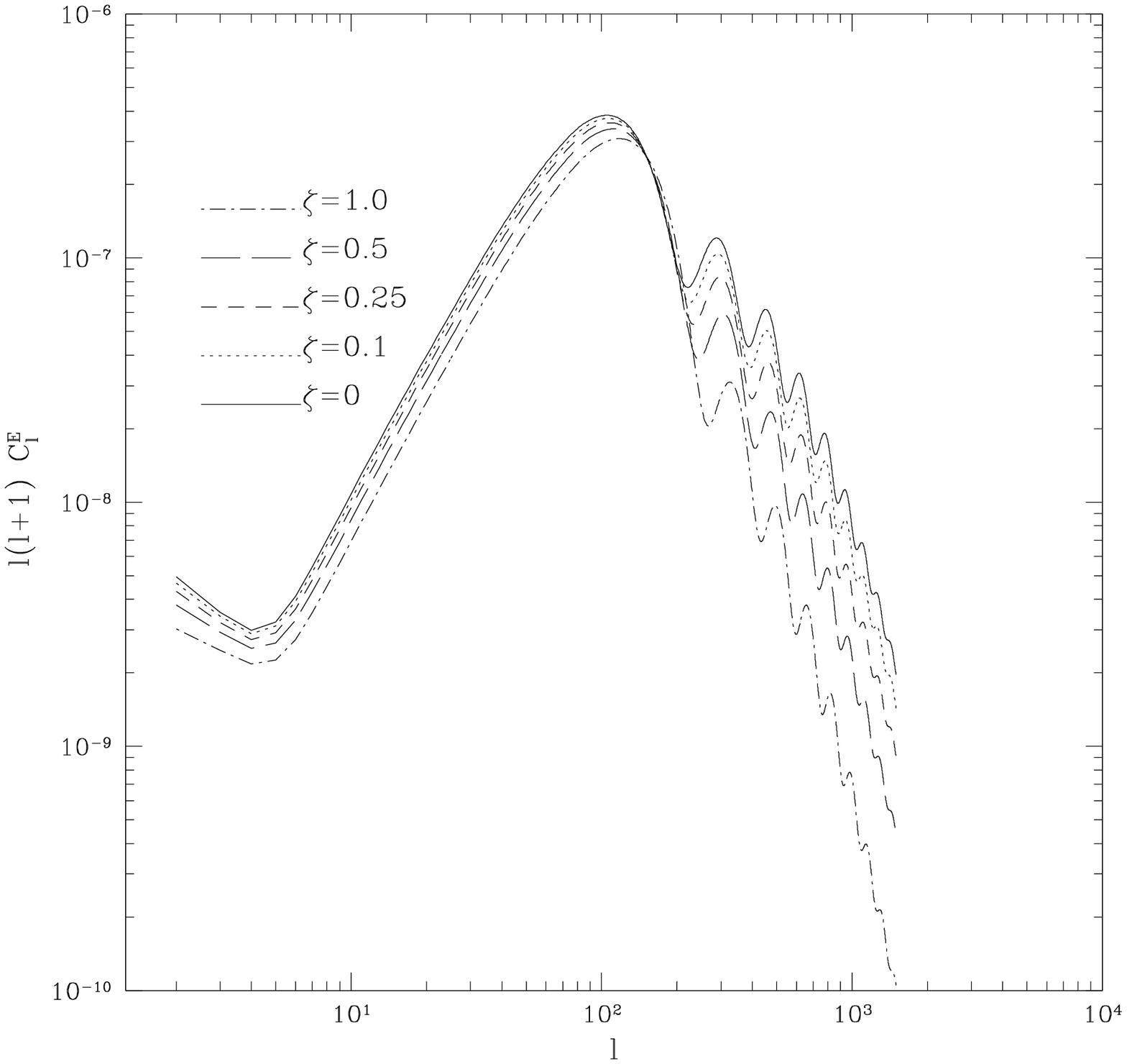}
\caption{ The electric polarization power spectrum for tensor
perturbations for the same braneworld models as in
Fig.~\ref{plot2b}. \label{plot3}}
\end{center}
\end{figure}

We use the initial conditions,
Eqs.~(\ref{e:hijini})--(\ref{e:elecini}), in a modified version of
the CAMB code to obtain the tensor temperature and polarization
power spectra. The temperature and electric polarization spectra
are shown in Figs.~\ref{plot2b} and \ref{plot3} for a
scale-invariant initial power spectrum. The normalisation is set
by the initial power in the gravity wave background. Figs
~\ref{plot2b} and \ref{plot3}, together with
Eqs.~(\ref{e:highenergy1})--(\ref{e:elecini}), are the main result of
this work, and we now discuss the physical conclusions
following from these results.     

\section{Discussion}

As expected, we find that the power spectra are insensitive to high-energy
effects, i.e. effectively independent of the brane tension $\lambda$ : the
$\zeta=0$ curve in Fig.~1 is indistinguishable from that of the
general relativity model (both power spectra are identical 
at the resolution of the plot). For the computations, we have used the
lowest value of the brane tension $\lambda=(100~{\rm  GeV})^4$,
 consistent with the tests of Newton's law.

There are three notable effects visible in Figs.~1 and 2, arising as
the physical consequences of our approximate model of the KK stress:
(i)~the power on large scales reduces with 
increasing KK parameter $\zeta$; (ii)~features in the spectrum
shift to smaller angular scales with increasing $\zeta$; and
(iii)~the power falls off more rapidly on small scales as $\zeta$
increases. Neglecting scattering effects, the shear is the only
source of linear tensor anisotropies (see e.g.\
Ref.~\cite{challinor4}). For $1\ll l < 60$ the dominant modes to
contribute to the temperature $C_l$s are those whose wavelengths
subtend an angle $\sim 1 / l$ when the shear first peaks (around
the time of Hubble crossing). The small suppression in the $C_l$s
on large scales with increasing $\zeta$ arises from the reduction
in the peak amplitude of the shear at Hubble entry [see
Eq.~(\ref{e:matterlong})], qualitatively interpreted as the loss
of energy in the 4D graviton modes to 5D KK modes.

Increasing $\zeta$ also has the effect of adding a small positive
phase shift to the oscillations in the shear on sub-Hubble scales,
as shown e.g.\ by Eq.~(\ref{e:mattershort}). The delay in the time
at which the shear first peaks leads to a small increase in the
maximum $l$ for which $l(l+1)C_l$ is approximately constant, as is
apparent in Fig.~\ref{plot2b}. The phase shift of the subsequent
peaks in the shear has the effect of shifting the peaks in the
tensor $C_l$s to the right. For $l > 60$ the main contribution to
the tensor anisotropies at a given scale is localized near last
scattering and comes from modes with wavenumber $k \sim l /
\tau_0$, where $\tau_0$ is the present conformal time. On these
scales the gravity waves have already entered the Hubble radius at
last scattering. Such modes are undergoing adiabatic damping by
the expansion and this results in the sharp decrease in the
anisotropies on small scales. Increasing the KK parameter $\zeta$
effectively produces more adiabatic damping and hence a sharper
fall off of power. The transition to a slower fall off in the
$C_l$s at $l \sim 200$ is due to the weaker dependence of the
amplitude of the shear on wavenumber at last scattering for modes
that have entered the Hubble radius during radiation
domination~\cite{starobinsky}. [The asymptotic expansion of
Eq.~(\ref{e:matterlong}) gives the shear amplitude $\propto k^{-(2
+ \zeta)}$ at fixed $\tau$, whereas for modes that were sub-Hubble
at matter-radiation equality Eq.~(\ref{e:mattershort}) gives the
amplitude $\propto k^{-(1 + \zeta/2)}$.]

Similar comments apply to the tensor electric polarization
$C^E_l$, shown in Fig.~\ref{plot3}. As with the temperature
anisotropies, we see the same shifting of features to the right
and increase in damping on small scales. Since polarization is
only generated at last scattering (except for the feature at very
low $l$ that arises from scattering at reionization, with an assumed 
optical depth $\tau_C=0.03$), the large-scale polarization is
suppressed, since the shear (and hence 
the temperature quadrupole at last scattering) is small for
super-Hubble modes. In matter domination the large-scale shear is
$\sigma_k = - k\tau / (5 + 2\zeta)$; the reduction in the
magnitude of the shear with increasing KK parameter $\zeta$ is
clearly visible in the large-angle polarization. The braneworld
modification to the tensor magnetic polarization $C^B_l$ has the
same qualitative features as in the electric case.

In principle, observations can constrain the KK parameter $\zeta$,
which controls the generation of 5D modes within our simplified
local approximation, Eq.~(\ref{e:Pansatz1}). The other braneworld
parameter $\lambda$, the brane tension, is not constrained within
our approximation. In practice, the tensor power spectra have not been
measured, and the prospect of useful data is still some way off.
What is more important is the theoretical task of improving on the
simplified local approximation we have introduced. This
approximation has allowed us to encode aspects of the qualitative
features of braneworld tensor anisotropies, which we expect to
survive in modified form within more realistic approximations.
However, a proper understanding of braneworld effects must
incorporate the nonlocal nature of the KK graviton modes, as
reflected in the general form of Eq.~$\eqref{e:soln}$. It is also
necessary to investigate the scalar anisotropies, which have a
dominant contribution to the measured power spectra. These may reveal new
braneworld imprints that are more amenable to observational testing.

\begin{acknowledgments}

B.L.\ thanks the Universities of Cape Town and Portsmouth for
hospitality, during which part of the work was done. B.L.\ thanks
Y.L.\ Loh, A.\ Lewis, A.\ Kahle, G.\ Chon and J. Weller for
assistance with various computational matters, P.\ Dunsby and C.
van de Bruck for helpful and critical discussions. B.L.\ is
supported by an Overseas Research Studentship, the Cambridge
Commonwealth Trust and the Lee Foundation, Singapore. A.C.\
acknowledges a PPARC Postdoctoral Fellowship.

\end{acknowledgments}

\end{document}